###################################### Latex

\documentclass[12pt]{article}

\begin{document}
\begin{flushright}
Stockholm\\ USITP 05-03\\ April 2003
\end{flushright}

\thispagestyle{empty}
\medskip
\begin{center}

{\large\bf Extra Gauge Symmetry for Extra Supersymmetry}

\vspace{10mm}
 A.A. Zheltukhin$^{\rm a,b}$ and D.V. Uvarov$^{\rm b} $
\end{center}
\begin{center}
$^{a}$ Institute of Theoretical Physics, University of Stockholm, SCFAB,\\
 SE-106 91 Stockholm, Sweden \\
$^{b}$ Kharkov Institute of Physics and Technology, 61108, Kharkov,  Ukraine
\vskip 15.mm
\end{center}
\begin{center}
{\bf Abstract}
\end{center}   
\vskip 5.mm

 It is shown that the extra supersymmetry of tensionless superstring and super $p$-brane is accompanied by the presence of new bosonic gauge symmetries. 
It permits to use composed coordinates encoding all physical degrees of freedom of the model and invariant under these gauge symmetries and the enhanced $\kappa$-symmetry. 
It is proved that the composed gauge invariant coordinates coincide with the componets of symplectic supertwistor realizing a linear representation of the hidden OSp(1,2M) symmetry of the super $p$-brane Lagrangian. A connection of the presented gauge symmetries with massless higher spin gauge theories and a symmetric phase of $M$/string-theory is discussed.

\vspace{10mm}

\section {Introduction}

Tensionless (super)strings have recently been discussed in the frame of massless higher spin field theory and AdS/CFT correspondence \cite{witten},\cite{bo},\cite{vas}. The point is the relation $g^{2}_{YM}N=(R^2/\alpha^{\prime})^2$ between the 't Hooft  coupling constant  $g^{2}_{YM}N$  and the rescaled string tension $R^{2}/\alpha^{\prime}$, where $R$ is the radius of 
$AdS_{5}\times S^5$ \footnote{The rescaled string tension as a perturbative parameter of string dynamics in curved spaces was previously considered in \cite {zhro}.}. This relation shows that the zero limit for the string tension $T= 1/\alpha^{\prime}$ leads to a free gauge theory. The conjecture was advanced that conformal ${\cal N}=4 $ SYM theory has to be dual to the theory of massless higher spin fields when $N$ is large.
Symplectic (super)symmetries  $OSp(1,2^p)$ play an important role in the formulation of the massless higher spin field dynamics and may be linearly realized in the generalized space-time described  by the real symmetric matrix $Y_{ab}\; (a,b=1,....,2^p)$  
and its superpartner  \cite{frons}, \cite{vas}. 
If $a,b$ are identified with the Majorana spinor indices the $Y_{ab}$ components can be treated as space-time coordinates added by tensor central charge (TCC) coordinates presented  by antisymmetric space-time tensors contracted with appropriate antisymmetrized products of 
$\gamma$-matrices.

It was assumed in \cite{vas} that higher spin gauge theory may be treated 
as a symmetric phase of $M$-theory. This assumption is supported by the observation  \cite{west} on a nonlinear character of the $OSp(1,64)$ supersymmetry realization in  
 $D=11$ supergravity which is the low-energy limit of $M$/string-theory, where higher spin exitations are massive. The consideration of superstring theory as a  spontaneously broken phase of higher spin gauge theory promises a new  basis 
for  understanding the microscopic structure of superbranes in terms of the 
spontaneous breaking parameters of higher spin gauge theory which correlates  with the topological approach \cite{agit}.
 
Thus, it is important to find the right kind of string/M-theory in terms of the new dynamical variables associated with orthosymplectic (super)symmetries and higher spins. This problem stimulates investigation of the superstring and superbrane dynamics in generalized spacetimes extended by the addition of TCC coordinates \cite{curt}, \cite{bersez}, \cite{zli}. 

Using the twistor-like approach \cite{guz},\cite{zli} the exactly solvable model of tensionless super $p$-brane preserving $3/4$ of the $D=4\, N=1$ supersymmetry was proposed in \cite{zuv}, which generalizes the superparticle model \cite{balu}, where preserving $3/4$ supersymmetry was earlier noted. This super $p$-brane saturates the exotic BPS state with the same symmetry whose existence  has been  proved in \cite{gght} applying the model independent analysis of the $D=4\, N=1$ superalgebra enlarged by tensor central charges.
The model \cite{zuv} preserves all its properties in the $D$-dimensional Minkowski space with  $D=2,3,4 \; mod(8)$, where the supersymmetric Cartan form and an auxiliary twistor-like Majorana spinor, defining the superbrane Lagrangian, are the same modulo the spinor's dimension.
 As a result, the tensionless super $p$-brane will saturate the exotic BPS state spontaneously breaking only one from  $M$ global supersymmetries, where $M$ is the Majorana spinor dimension. An interpretation of this tensionless super p-brane as the BPS state of $M$-theory was recently discussed in \cite{band}. 
An important property of the tensionless super $p$-brane \cite{zuv} and  the
appropriate  BPS state is the linear character of the $OSp(1,2M)$ symmetry realization. The linear realization is a result of the transition to a new 
  supertwistor variable previously considered  in the superparticle dynamics \cite{ferb},\cite{shir},\cite{balu}. The symplectic supertwistor $Y^\Lambda$ \cite{zuv} encodes all physical degrees of freedom contained in $Y_{ab}$ and the Majorana spinor $\theta^a$ which describe the original formulation of the model. 
This reduction of the initial variable number points out on the presence of hidden bosonic gauge symmetries strongly correlating with the extra $\kappa$-symmetry. A fine tuning of these symmetries results in the minimal spontaneous breaking of the $N=1$ global supersymmetry and manifests itself by the $OSp(1,2M)$ symmetry appearance similarly to the superparticle case \cite{balu}, \cite{balus}, where the first class constraints have signaled about the hidden bosonic symmetry presence. 

Therefore, the problem  appears to find hidden gauge symmetries responsible for the reduction of the bosonic degrees of freedom described by the space-time and TCC coordinates. Let us remind that the reduction of the fermionic degrees of freedom of the model is provided by its enhanced $\kappa$-symmetry \cite{zuv}.
The required bosonic symmetries should match the enhanced $\kappa$-symmetry and their description is important for understanding the structure of a symmetric phase of string/$M$-theory \cite{vas} and local symmetries compatible with the BPS states characterized by  extra supersymmetry \cite{gah}.

In this letter we present a set of new bosonic gauge symmetries matched with the enhanced $\kappa$-symmetry of the tensionless super $p$-brane. We show that these gauge symmetries result in the invariant reduction of the bosonic gauge degrees of freedom described by the original symplectic spin-tensor $Y_{ab}$. The invariant character of the reduction means that the symplectic supertwistor $Y^\Lambda$ encoding all physical degrees of freedom and providing the $OSp(1,2M)$ symmetry of the superbrane Lagrangian is gauge invariant under the transformations of the bosonic and enhanced $\kappa$ symmetries. It should be noted that the bosonic symmetries include the Weyl gauge symmetry which  correlates with a global space-time conformal symmetry of the tensionless string action as it was  noted in \cite{ilst}.

\section {Tensionless super $p$-brane with enhanced supersymmetry}

The inclusion of TCC in the algebra of $ N=1$ supersymmetry generalizes anticommutator of the Majorana supercharges $Q_a$ to the form \cite{dafre}, \cite{vanvan}
\begin{equation}\label{1}
\{ Q_a ,Q_b \}=(\gamma^m\,C^{-1})_{ab}P_{m}+i(\gamma^{mn}\,C^{-1})_{ab}Z_{mn}+
(\gamma^{mnl}\,C^{-1})_{ab}Z_{mnl}+.....
\end{equation} 
The r.h.s. of Eq. (\ref{1}) contains antisymmetrized products of $\gamma$-matrices multiplied by the TCC $Z_{mn...}$ which are antisymmetric Lorentz tensors in the Minkowski space-time commuting with $Q_a$.
The real antisymmetric parameters $z_{mn...}$ corresponding to $Z_{mn...}$ are known as TCC coordinates which may be presented in the equivalent spinor form 
\begin{equation}\label{2}
z_{ab}=iz_{mn}(\gamma^{mn}\,C^{-1})_{ab}+z_{mnl}(\gamma^{mnl}\,C^{-1})_{ab}+....
\end{equation} 
by analogy with the spinor representation of  the space-time coordinates
$x_{m}$
\begin{equation}\label{3}
x_{ab}=x_{m}(\gamma^{m}\,C^{-1})_{ab}.
\end{equation} 
The coordinates $z_{ab}$ and $x_{ab}$ may be treated as the components of
the real symmetric spin-tensor
\begin{equation}\label{4}
Y_{ab}\equiv x_{ab} + z_{ab}
\end{equation} 
which can be identified with the symmetric matrix of generalized symplectic coordinates 
considered in \cite{frons}, \cite{vas}.
The pairs $(Y_{ab}, \theta_a)$ unifying  $Y_{ab}$ and the Majorana spinor $\theta_a$ form a generalized superspace invariant under the $N=1$ global supersymmetry 
\begin{equation}\label{5} 
\delta_\varepsilon\theta_{a}=\varepsilon_a ,\quad 
\delta_\varepsilon Y_{ab}=2i(\theta_a\varepsilon_b +\theta_b\varepsilon_a ).
\end{equation} 
The differential one-forms $W_a$ and  $W_{ab}$ of the generalized superspace 
\begin{equation}\label{6}
W_a=d\theta_a,\quad W_{ab}=dY_{ab}-2i(d\theta_a\theta_b + d\theta_b\theta_a )
\end{equation}
are the Cartan forms invariant under (\ref{5}) which have been used to construct the exactly solvable model \cite{zuv} 
\begin{equation}\label{7}
S_p=\frac{1}{2}\int d\tau d^{\,p}\sigma \rho^\mu (U_a W_\mu^{ab} U_b),
\end{equation} 
of tensionless super $p$-brane ($p=1,2,3,...)$ with extra $\kappa$-symmetry. 
The action $S_p$ includes an auxiliary Majorana spinor $U_a $ \cite{volzh} and the world-volume density $\rho^\mu$ \cite{banzh} which are invariants of the $N=1$ supersymmetry. 
The spinor $U_a $ parametrizes the light-like density of the brane momentum. The action $S_p$ is also invariant under the transformations of the enhanced $\kappa-$symmetry 
\begin{eqnarray}\label{8}
\delta_\kappa\theta_a=\kappa_a  ,\quad
\delta_\kappa Y_{ab}=-2i(\theta_a\kappa_b + \theta_b\kappa_a ), \nonumber\\ 
  \delta_\kappa U_{a}=0, \quad \delta_\kappa \rho^\mu=0,
\end{eqnarray} 
where the parameter $\kappa$ is restricted by the only one real condition
\begin{equation}\label{9}
\kappa^{a} U_{a}=0.
\end{equation} 
As a result, this model preserves $(\frac{M-1}{M})$ fraction of the $N=1$ supersymmetry, where $M$ is the dimension of the correspondent Majorana spinor. 
The model yields a pure static general solution for the Goldstone fermion $\tilde\eta$ defined by the Lorentz invariant projection
 \begin{equation}\label{10}
\tilde\eta=-2i (U^{a}\theta_a)
\end{equation} 
encoding the spontaneously broken component of the supersymmetry. 
The exact solvability of the model implies the presence of hidden local symmetries which is revealed by the change of variables 
\begin{equation}\label{11} 
 i\tilde Y_a= Y_{ab}U_{}^b - \tilde \eta \theta_a
\end{equation} 
 introducing the Majorana spinor $\tilde Y_a$ as a new variable substituted for $ Y_{ab}$.
In terms of the new spinor variable  the action (\ref{7}) transforms to the form
\begin{equation}\label{12}
S_p=\frac{i}{2}\int d\tau d^{\,p}\sigma
\,\rho^\mu\{[(U^a\partial_\mu \tilde Y_a)
-(\partial_\mu U^a \tilde Y_a)] - 
\tilde\eta \partial_{\mu}\tilde\eta \}.
\end{equation} 
The action (\ref{12}) is the component representation of the $OSp(1,2M)$ invariant action  
  \begin{equation}\label{13} 
S_p=\frac{1}{2}\int d\tau d^{\,p}\sigma\,\rho^\mu 
\partial_{\mu}Y^{\Lambda} G_{\Lambda\Xi}Y^{\Xi}
\end{equation} 
in which the real $OSp(1,2M)$ supertwistor $Y^{\Lambda}=( iU^{a}, \tilde Y_a, \tilde\eta )$ and invariant supersymplectic metric $G_{\Lambda\Xi}= (-1)^{\Lambda\Xi+1}{}G_{\Xi\Lambda}$ 
have been used. The transition  from the representation (\ref{7}) to the supertwistor representation (\ref{12}) (or (\ref{13})) is accompanied by the reduction of some original variables both in the fermionic and bosonic sectors. It means that the super $p$-brane Lagrangian is singular due to the presence of hidden gauge symmetries. The enhanced $\kappa$-symmetry (\ref{8}) is responsible for the reduction of the $(M-1)$ of the $M$ components of the Majorana spinor $\theta_a$ and one remaining fermionic variable $\tilde \eta $ (\ref{10}) proved to be invariant under the $\kappa$-symmetry transformations (\ref{8}),(\ref{9}). The invariance of the Goldstone fermion $\tilde \eta $ proves that  $(M-1)$ fermionic gauge degrees of freedom have been reduced without gauge fixing. The question then appears about hidden gauge symmetries responsible for the reduction of the bosonic gauge degrees of freedom contained in $ Y_{ab}$. 
    We shall define these symmetries in the next Section.

\section {Gauge symmetries matched with extra $\kappa$-symmetry}
 
First of all we note that both of the representations (\ref{7}) and 
(\ref{12}) are invariant under the local Weyl symmetry including one real parameter $\Lambda(\tau,\vec\sigma)$
\begin{equation}\label{14}
\rho'^{\mu}=e^{-2\Lambda}\rho^\mu, \;  U'_a=e^\Lambda U_a, \; \theta'^{a}=\theta^{a}, \;  Y'_{ab}=Y_{ab}.
\end{equation}
 The transformations  (\ref{14}) imply that $x'_{ab}=x_{ab},\; z'_{ab}=z_{ab}$, but $\tilde Y_{a}$ and the supertwistor $Y'^\Lambda$ are not invariant under the Weyl transformations
\begin{equation}\label{15}
\tilde Y'_{a}=e^{\Lambda}\tilde Y_{a}, \;  Y'^\Sigma=e^{\Lambda}Y^\Sigma.
\end{equation}
The invariant character of $Y_{ab}$ (\ref{14}) means that the Weyl symmetry does not participate in the discusssed reduction of the bosonic coordinates and other gauge symmetries should be found. 
To connect the above-mentioned gauge symmetries with the $OSp(1,8)$ symmetry of the $4d$ higher spin theory \cite{vas} and the results \cite{gght} we present here a detailed analysis of the $D=4 \, N=1$ supersymmetry. The generalization of these results to  the higher dimensions $D=2,3,4 \; mod (8)$ will be clear from this analysis.

In the $4d$ case the action $S_p$  (\ref{7}) acquires the form
\begin{equation}\label{16}
S_p=\frac12\int d\tau d^p\sigma\ \rho^\mu 
\left(2u^\alpha\omega_{\mu\alpha\dot\alpha}\bar 
u^{\dot\alpha}+u^\alpha\omega_{\mu\alpha\beta}u^\beta+\bar 
u^{\dot\alpha}\bar\omega_{\mu\dot\alpha\dot\beta}\bar u^{\dot\beta}\right),
\end{equation}
where the supersymmetric one-forms $\omega_{\mu\alpha\dot\alpha}$ and 
$\omega_{\mu\alpha\beta}$ in the Weyl basis are
\begin{eqnarray}\label{17}
 \nonumber 
\omega_{\mu\alpha\dot\alpha}=\partial_\mu 
x_{\alpha\dot\alpha}+2i(\partial_\mu\theta_\alpha\bar\theta_{\dot\alpha}+\partial_\mu\bar\theta_{\dot\alpha}\theta_\alpha),\\ 
\nonumber 
\omega_{\mu\alpha\beta}=-\partial_\mu 
z_{\alpha\beta}-2i(\partial_\mu\theta_\alpha\theta_\beta+\partial_\mu\theta_\beta\theta_\alpha),\\  
\bar\omega_{\mu\dot\alpha\dot\beta}=-\partial_\mu\bar 
z_{\dot\alpha\dot\beta}-2i(\partial_\mu\bar\theta_{\dot\alpha}\bar\theta_{\dot\beta}+\partial_\mu\bar\theta_{\dot\beta}\bar\theta_{\dot\alpha}).
\end{eqnarray}
For the search of hidden gauge symmetries it is efficient to introduce a basis in the spinor space of the model. To this end a linearly independent local Weyl spinor $v^\alpha$ 
may be added to $u^\alpha$.  Then, without loss of generality, the Weyl spinors $u^\alpha$ and $v^\alpha$ attached to the brane worldvolume may be identified with the local Neuman-Penrose dyad \cite{newpen} defined by the well known relations
\begin{equation}\label{18}
u^\alpha u_\alpha=0,\quad 
v^\alpha v_\alpha=0,\quad 
u^\alpha v_\alpha\equiv u^\alpha\varepsilon_{\alpha\beta}v^\beta=1
\end{equation}
and their complex conjugate. The scalar products  (\ref{18}) will be invariants of the Weyl transformations (\ref{14}) if the transformation  $V'_a=e^{-\Lambda} V_a$ is taken into account. The Majorana bispinors $U_a(\tau,\vec\sigma)={u_\alpha\choose \bar u^{\dot\alpha}}$, $V_a(\tau,\vec\sigma)={v_\alpha\choose \bar v^{\dot\alpha}}$, \,$(\gamma^5U)_a$ and $(\gamma^5V)_a$ \cite{zuv} will respectively form a basis in the Majorana bispinor space.

The first of the desired gauge symmetries which transforms only $x_m$ is defined as 
\begin{equation}\label{19}
\delta_{X} x_{\alpha\dot\alpha}=\epsilon_{X}u_\alpha\bar u_{\dot\alpha}
\end{equation}
 and it is a local symmetry of the action $S_p$ (\ref{16}), due to the relation $u^\alpha u_\alpha=0$.
We shall call this one-parametric real transformation as $X$-shift. It shifts $x_m$ by the light-like $4$-vector $(u\sigma_m\bar u)$
\begin{equation}\label{20}
\delta_{X}x_{m}=-\frac12\epsilon_{X}(u\sigma_m\bar u).
\end{equation}
The change $(u\sigma_m\bar u)\rightarrow (\bar U\gamma_mU)$ in (\ref{20}) lifts the  $X$-shift to the bispinor representation and shows that it is also the symmetry of the high dimensional action (\ref{7}) due to the relation $\quad(U^aU_a)$=0 satisfiable for the Majorana bispinors. 

The next gauge symmetry of $S_p$ (\ref{16}) transforms only the TCC coordinates $z_{\alpha\beta}$ 
\begin{equation}\label{21}
\delta_{T} z_{\alpha\beta}=\epsilon_{T} u_\alpha u_\beta,\quad
\delta_{T}\bar z_{\dot\alpha\dot\beta}=\bar\epsilon_{T}\bar u_{\dot\alpha} \bar u_{\dot\beta} \;.
\end{equation}
It includes one complex parameter $\epsilon_{T}$ and we shall call it as $T$-shift. Note that the local spin-tensor $u_\alpha u_\beta$ in (\ref{21}) has zero norm, i.e. $(u_\alpha u_\beta)(u^\alpha u^\beta)$=0, and defines a set of local null planes attached to the super $p$-brane world-volume. Therefore, the $T$-shifts are a generalization of the $X$-shifts (\ref{19}), 
described by the field of null vectors, to the shifts defined by the field of null bivectors, as it can be seen from the representation (\ref{21}) in the tensor form 
\begin{equation}\label{22}
\delta_T z_{mn}=-\frac{i}{4}[\epsilon^{(R)}_T(u\sigma_{mn}u+\bar 
u\tilde\sigma_{mn}\bar u)+\epsilon^{(I)}_T(u\sigma_{mn}u-\bar 
u\tilde\sigma_{mn}\bar u)],
\end{equation}
where $\epsilon^{(R)}$ and $\epsilon^{(I)}$ are real parameters
\begin{equation}\label{23}
\epsilon^{(R)}_T=\frac12(\epsilon_{T}+\bar\epsilon_{T}),\quad
\epsilon^{(I)}_T=\frac{1}{2i}(\epsilon_{T}-\bar\epsilon_{T}). 
\end{equation}   
In terms of the Majorana bispinor $U_a$ the transformation (\ref{22}) is presented as
\begin{equation}\label{24}
\delta_T z_{mn}=\frac{i}{8}[\epsilon^{(R)}_T(\bar 
U\gamma_{mn}U)+\epsilon^{(I)}_T(\bar U\gamma_{mn}\gamma_5 U)],
\end{equation}
 where  the bivectors $(\bar U\gamma_{mn}U)$ and $(\bar U\gamma_{mn}\gamma_5U)$ belong to 
the field of null or isotropic bivectors \cite{guz} defined  by the conditions 
\begin{equation}\label{25}
(\bar U\gamma_{mn}U)^2=0,\quad(\bar U\gamma_{mn}\gamma_5U)^2=0.
\end{equation}
and are interpreted as the egenvalues of the generalized TCC $Z_{mn}$ (\ref{1}). We see that the local translations of the TCC coordinates $z_{mn}$ by the null bivectors (\ref{25}) produce a new type of gauge symmetry due to which the TCC coordinates $z_{mn}$ proved to be defined modulo the shift by the null bivectors. In the higher dimensional Minkowski spacetimes additional multivector translations presented  by the bilinear covariants similar to $(\bar U\gamma_{mn...l}U)$  will appear as admissible gauge symmetries of the action (\ref{7}).

To continue the description of the next gauge symmetries we note that 
the dyad space (\ref{18}) is symmetric under the $u\leftrightarrow v$ spinor permutation which transforms the null spin-tensors  $u_\alpha u_\beta$ 
(\ref{21}) to the null spin-tensors $v_\alpha v_\beta$. But, it is not a symmetry of the action (\ref{16}).
 However, the local shift of $ z_{\alpha\beta}$ by the null tensor $ v_\alpha v_\beta$ similar to the null shift (\ref{21}) may be compensated if simultaneous shift of $ x_{\alpha\dot\alpha}$ by the correspondent null vector 
$v_{\alpha} \bar v_{\dot\alpha}$ similar to (\ref{19})
will be added.
As a result, we find a new one-parametric gauge symmetry of $S_p$ (\ref{16})
\begin{equation}\label{26}
\delta_{\tilde T_R}x_{\alpha\dot\alpha}=\epsilon_{\tilde 
T_R}v_\alpha\bar v_{\dot\alpha},\quad
\delta_{\tilde T_R}z_{\alpha\beta}=\epsilon_{\tilde T_R}v_\alpha 
v_\beta,
\quad\delta_{\tilde T_R}\bar 
z_{\dot\alpha\dot\beta}=\epsilon_{\tilde T_R}\bar v_{\dot\alpha}\bar v_{\dot\beta},
\end{equation}
because the correspondent variation of $S_p$ (\ref{16})
\begin{equation}\label{27}
\delta_{\tilde T_R}S_p=\int d\tau 
d^p\sigma\rho^\mu\partial_\mu\epsilon_{\tilde T_R}[(u^\alpha 
v_\alpha)^2-1]=0
\end{equation}
equals to zero due to (\ref{18}). Let us  call the transformation (\ref{26}) as $\tilde T_R$-shift.
 
The local space-like vectors $m^{(+)}_{\alpha\dot\alpha}$ and $m^{(-)}_{\alpha\dot\alpha}$
\begin{equation}\label{28}
m^{(+)}_{\alpha\dot\alpha}=u_\alpha\bar v_{\dot\alpha}+v_\alpha\bar 
u_{\dot\alpha},\quad 
m^{(-)}_{\alpha\dot\alpha}=i(u_\alpha\bar 
v_{\dot\alpha}-v_\alpha\bar u_{\dot\alpha}),\quad 
n^{(+)}_{\alpha\dot\alpha}=u_\alpha\bar u_{\dot\alpha},\quad 
n^{(-)}_{\alpha\dot\alpha}=v_\alpha\bar v_{\dot\alpha},
\end{equation}
 orthogonal to the real local null vectors  $u_\alpha\bar u_{\dot\alpha}$ (\ref{19}) and $v_\alpha\bar v_{\dot\alpha}$  (\ref{26}), form the local tetrade attached to the superbrane worldvolume.  

The local shifts of the $x$-coordinates in the transverse directions 
$m^{(\pm)}_{\alpha\dot\alpha}$
\begin{equation}\label{29}
\delta_{\Phi^{(+)}}x_{\alpha\dot\alpha}=\epsilon_{\Phi^{(+)}}m^{(+)}_{\alpha\dot\alpha},\quad\delta_{\Phi^{(-)}}x_{\alpha\dot\alpha}=\epsilon_{\Phi^{(-)}}m^{(-)}_{\alpha\dot\alpha},
\end{equation}
which we shall call $\Phi^{(\pm)}$-shifts, change $S_p$ (\ref{16})
\begin{eqnarray}\label{30}
\nonumber
\delta_{\Phi^{(+)}}S=\int d\tau 
d^p\sigma\rho^\mu\epsilon_{\Phi^{(+)}}(u^\alpha\partial_\mu u_\alpha+\bar 
u^{\dot\alpha}\partial_\mu\bar u_{\dot\alpha}),\\
\delta_{\Phi^{(-)}}S=i\int d\tau 
d^p\sigma\rho^\mu\epsilon_{\Phi^{(-)}}(u^\alpha\partial_\mu u_\alpha-\bar 
u^{\dot\alpha}\partial_\mu\bar u_{\dot\alpha}).
\end{eqnarray}
However, just as in the previous case the variations (\ref{30}) are exactly compensated by the correspondent shifts of the TCC coordinates $z_{\alpha\beta}$
\begin{eqnarray}\label{31}
\nonumber
\delta_{\Phi^{(+)}}z_{\alpha\beta}=2\epsilon_{\Phi^{(+)}}
u_{\{\alpha}v_{\beta\}} ,\quad
\delta_{\Phi^{(+)}}\bar 
z_{\dot\alpha\dot\beta}=2\epsilon_{\Phi^{(+)}}\bar u_{\{\dot\alpha}\bar 
v_{\dot\beta\}};\\
\delta_{\Phi^{(-)}}z_{\alpha\beta}=2i\epsilon_{\Phi^{(-)}}u_{\{\alpha}v_{\beta\}},\quad 
\delta_{\Phi^{(-)}}\bar 
z_{\dot\alpha\dot\beta}=-2i\epsilon_{\Phi^{(-)}}\bar u_{\{\dot\alpha}\bar 
v_{\dot\beta\}},
\end{eqnarray}
where the symmetrized production $u_{\{\alpha}v_{\beta\}}\equiv \frac{1}{2}(u_{\alpha}v_{\beta}+u_{\beta}v_{\alpha})$ was introduced.

Thus, we found six real bosonic gauge symmetries of the action (\ref{7}) in the  $4d$ Minkowski space-time generated by the $X,\; T, \; \bar T,\; \tilde T_R, \;  \Phi^{(+)}$ and 
$\Phi^{(-)}$ which form six parametric abelian group of translations in the $10d$ symplectic subspace of the $11d$ symplectic superspace. 
 In the next Section we shall show  that these gauge symmetries are responsible for the invariant reduction of six bosonic gauge degrees of freedom contained in $Y_{ab}$ (\ref{4}).

\section {The gauge invariance of the supertwistor $Y^\Lambda$}

Here we show that the supertwistor $Y^\Lambda$ is invariant under the above presented 
bosonic gauge symmetries and the enhanced $\kappa$-symmetry.
The invariance of $Y^\Lambda$ will prove the gauge invariant character of the considered reduction of bosonic and fermionic degrees of freedom. 
To this end let us consider the transformation properties of the supertwistor $Y^\Lambda=( iU^{a}, \tilde Y_a, \tilde\eta )$ under the fermionic $\kappa$-symmetry (\ref{8}), (\ref{9}) and bosonic symmetries described by the $X$-shifts (\ref{19}), $T$-shifts (\ref{21}), $\tilde T_R$-shifts (\ref{26}) and the $\Phi^{(\pm)} $-shifts (\ref{29}), (\ref{31}). 

Then we find that invariant character of the $U^{a}$ and $\tilde\eta $ follows from the definitions of the transformation rules of the enhanced $\kappa$-symmetry and the bosonic symmetries. 

Using this observation one can present the gauge transformations of the remaining supertwistor component  $i\tilde Y_a$ as
\begin{equation}\label{32}
i\delta\tilde Y_a=\delta Y_{ab}U^b-\tilde\eta\delta\theta_a.
\end{equation}
The substitution of the $\kappa$-symmetry transformations (\ref{8}) in (\ref{32}) together 
with using  the relations (\ref{9}) and (\ref{10}) yields the required result
\begin{equation}\label{33}
\delta_\kappa\tilde Y_a=0.
\end{equation}

Taking into account the invariance of $\theta_a$ under the bosonic gauge symmetries one can  simplify the variation  (\ref{32}) to the form
\begin{equation}\label{34}
i\delta\tilde Y_a=\delta Y_{ab}U^b ={-\delta z_{\alpha\beta}u^\beta+\delta 
x_{\alpha\dot\alpha}\bar u^{\dot\alpha}\choose\delta\tilde 
x^{\dot\alpha\alpha}u_\alpha-\delta\bar z^{\dot\alpha\dot\beta}\bar 
u_{\dot\beta}}.
\end{equation}
The substitution of the  $X$-shifts (\ref{19}) and  $T$-shifts (\ref{21}) into (\ref{34}) results in the relations
\begin{equation}\label{35}
\delta_{X} x_{\alpha\dot\alpha}\bar 
u^{\dot\alpha}=\epsilon_{X}u_\alpha(\bar u_{\dot\alpha}\bar 
u^{\dot\alpha})=0,\quad\delta_{T} 
z_{\alpha\beta}u^{\beta}=\epsilon_{T}u_\alpha(u_{\beta}u^{\beta})=0
\end{equation}
and their complex conjugate which prove the invariance of $\tilde Y_a$ under these shifts.

The invariance $\tilde Y_a$ under the $\tilde T_R$-shifts (\ref{26}) follows from the cancellation of $x$ and $z$ contributions given  by  
\begin{equation}\label{36}
\delta_{\tilde T_R} x_{\alpha\dot\alpha}\bar u^{\dot\alpha}-\delta_{\tilde T_R} 
z_{\alpha\beta}u^\beta=\epsilon_{\tilde T_R} v_\alpha-\epsilon_{\tilde T_R} v_\alpha=0.
\end{equation}

The analogous cancellations take place between the $x$ and $z$ contributions 
\begin{eqnarray}\label{37}
\nonumber
\delta_{\Phi^{(+)}} x_{\alpha\dot\alpha}\bar u^{\dot\alpha}-\delta_{\Phi^{(+)}} 
z_{\alpha\beta}u^\beta=\epsilon_{\Phi^{(+)}} [m^{(+)}_{\alpha\dot\alpha}\bar 
u^{\dot\alpha}-2u_{\{\alpha}v_{\beta\}}u^\beta]=0,\\ 
\delta_{\Phi^{(-)}} 
x_{\alpha\dot\alpha}\bar u^{\dot\alpha}-\delta_{\Phi^{(-)}} 
z_{\alpha\beta}u^\beta=\epsilon_{\Phi^{(-)}} [m^{(-)}_{\alpha\dot\alpha}\bar 
u^{\dot\alpha}-2 iu_{\{\alpha}v_{\beta\}}u^\beta]=0
\end{eqnarray}
and their complex conjugate generated by the $\Phi^{(\pm)} $-shifts (\ref{29}), (\ref{31}).

It completes the proof of the invariance of $Y^\Lambda$ under the discussed six bosonic and three fermionic gauge symmetries.

\section {Conclusion}

Symmetries of tensionless superstring and superbrane with extra supersymmetry were studied. It was shown that the $\kappa$-symmetry enhancement is accompanied by the appearance of bosonic gauge symmetries including the Weyl transformation and the local abelian translations of the space-time and TCC coordinates. In the case of $ D=4 \, N=1$ supersymmetry the translations are presented by the local vectors and bivectors constructed from the components of an auxiliary spinor field parametrizing the momentum density of the brane. In the high dimensional spaces gauge multivector translations of TCC coordinates will appear. Due to these gauge symmetries the original brane's coordinates are defined modulo these gauge translations resulting in the appearance of new composed coordinates encoding the physical degrees of freedom contained in the original coordinates. The new  variables are unified in the componets of a symplectic supertwistor realizing linear representation of the OSp(1, 2M) symmetry. We proved that this supertwistor is an invariant of the gauge translations and the enhanced $\kappa$-symmetry. 
So, the linearly realizible symplectic supersymmetries describing massless higher spin gauge theories appear as a result of a fine tuning of the set of local and global symmetries of the 
Lagrangians of tensionless superstring and superbrane. The tensionless objects are connected with the high energy limit $E\gg M_{Planck}$ of the string theory, where masses of all particles are negligible and a hidden large symmetry takes shape \cite{grom}. 
It hints that symplectic supertwistor describing tensionless strings and branes may be found relevant variable for the description of a symmetric phase of 
$M$-theory and quantum field string theory at the high energy scale.

 \section {Acknowledgements}

A.Z. thanks Fysikum at the Stockholm University for the kind hospitality and Ingemar Bengtsson for the useful discussion. The work was partially supported by the grant of the Royal Swedish Academy of Sciences and Ukrainian SFFR project 02.07/276.


\begin{thebibliography}{99}

\bibitem{witten}
 E. Witten, unpublished (see http://theory.caltech.edu/jhs60/witten/1.html).
\bibitem{bo}
B. Sundborg, Nucl. Phys. (Proc. Suppl.) B 102/101 (2001) 113. 
\bibitem{vas}  
M.A. Vasiliev,  Phys. Rev. D 66 (2002) 066006;\\ Russ. Phys. J. 45 (2002) 670; Izv. Vuz. Fiz. 2002 N7 (2002) 23. 
 \bibitem{zhro}
A.A. Zheltukhin, CQG 13 (1996) 2357;\\  S.N. Roshchupkin and A.A. Zheltukhin,
Nucl. Phys. B 543 (1999) 365.
\bibitem{frons}
C. Fronsdal, Masslesss particles, orthosymplectic symmetry and another type of Kaluza-Klein theory, Preprint UCLA/85/TEP/10, in Essays on supersymmetry, \\Reidel, 1986 (Mathematical Physics Studies, v.8).
\bibitem{west}
P.C. West, JHEP 08 (2000) 007.
\bibitem{agit}
J.A. de Azcarraga, J.P. Gauntlett, J.M. Izquierdo and  P.K. Townsend,\\
 Phys. Rev. Lett. 63 (1989) 2443.
\bibitem{curt}
T. Curtright, Phys. Rev. Lett. 60 (1988) 393.
\bibitem{bersez}
E. Bergshoeff and E. Sezgin,
Phys. Lett. B 392 (1995) 256.
\bibitem{zli}
 A.A. Zheltukhin and  U. Lindstr\"om,  Nucl. Phys. (Proc. Suppl.) B 102/101 (2001) 126; 
JHEP 01 (2002) 034;
\bibitem{guz}
 O.E. Gusev and  A.A. Zheltukhin, JETP Lett. 64 (1996) 487.
\bibitem{zuv}
 A. A. Zheltukhin and D.V. Uvarov, Phys. Lett. B 545 (2002) 183;\\
JHEP 08 (2002) 008.
\bibitem{balu} 
I. Bandos and J. Lukierski, Mod. Phys. Lett. A 14 (1999) 1257.
\bibitem{gght}
J.P. Gauntlett, G. Gibbons, C.M. Hull and P.K. Townsend, \\
Comm. Math. Phys. 216 (2001) 431.
\bibitem{band}
 I.A. Bandos,  Phys. Lett. B 558 (2003) 197.
\bibitem{ferb}
A. Ferber, Nucl. Phys. B 132 1978 55.
\bibitem{shir}
T. Shirafuji, Progr. Theor. Phys. 70 (1983) 18.
\bibitem{balus}
I. Bandos, J. Lukierski and D. Sorokin, Phys. Rev. D 61 (2000) 045002;\\
I. Bandos, J. Lukierski, C. Preitschopf and D. Sorokin, Phys. Rev. D 61 (2000) 065009.
\bibitem{gah}
J.P. Gauntlett and C.M. Hull, JHEP 01 (2000) 004.
\bibitem{ilst}
I. Izberg,  U. Lindstr\"om  and  B. Sundborg, 
Phys. Lett. B 293 (1992) 321;\\
I. Izberg,  U. Lindstr\"om,  B. Sundborg and G. Theodoridis,
Nucl. Phys. B 411 (1994) 122.
\bibitem{dafre}
 R. D'Auria and P. Fre, Nucl. Phys. B 201 (1982) 101;  erratum Nucl. Phys. B 206 (1982) 496.
 \bibitem{vanvan}
J. van Holten and A. van Proyen, J. Physics A 15 (1982) 101.
\bibitem{volzh}
D.V. Volkov and A.A. Zheltukhin,
JETP Lett. 48 (1988) 63; \\Lett. Math. Phys. 17 (1989) 141.
\bibitem{banzh}
 I.A. Bandos and  A.A. Zheltukhin, Fortschr. Phys. 41 (1993) 619.
\bibitem{newpen} R. Penrose and M.A.H. Mac Callum, Phys. Rep. 6 (1972) 241.
\bibitem{grom}
D.J. Gross and P.F. Mende,  Phys. Lett. B 197 (1987) 129;\\
D.J. Gross, unpublished (see http://theory.caltech.edu/jhs60/gross/13.html).

\end{thebibliography}
\end{document}